\documentclass{article}

\usepackage{PRIMEarxiv}

\usepackage[utf8]{inputenc} 
\usepackage[T1]{fontenc}    
\usepackage[colorlinks=true, linkcolor=black, citecolor=black, urlcolor=black]{hyperref}
\usepackage{url}            
\usepackage{booktabs}       
\usepackage{amsfonts}       
\usepackage{nicefrac}       
\usepackage{microtype}      
\usepackage{lipsum}
\usepackage{fancyhdr}       
\usepackage{graphicx}       
\graphicspath{{media/}}     

\title{

Rescue Operators' Perspectives on KIRETT Wearable Technology: A Qualitative Study

\thanks{\textit{
2024 IEEE International Conference on Systems, Man, and Cybernetics (SMC) | 978-1-6654-1020-5/24/\$31.00 ©2024 IEEE | DOI: 10.1109/SMC54092.2024.10831087
}}}

\author{
  Mubaris Nadeem, Johannes Zenkert, Lisa Bender, Christian Weber, Madjid Fathi \\
  Institute for Knowledge-Based Systems and Knowledge Management \\
  University of Siegen\\
  Hoelderlinstrasse 3, 57068 Siegen, Germany\\
  \texttt{\{Mubaris Nadeem\}mubaris.nadeem@uni-siegen.de} \\
}

\begin{document}
\maketitle

\begin{abstract}

In emergencies, treatment needs to be fast, accurate and patient-specific. For instance, in emergency scenarios, obstacles like treatment environments and medical difficulties can lead to bad outcomes for patients. Additionally, a drastic change of health vitals can force paramedics to shift to a different treatment in the ongoing treatment of the patient in order to save a patient's life. The KIRETT (engl.: ‘Artificial intelligence in rescue operations’) demonstrator is developed to provide a rescue operator with a wrist-worn device, enabling treatment recommendation (with the help of knowledge graph) with situation detection models to improve the emergency treatment of a patient. This paper aims to provide a qualitative evaluation of the 2-days testing in the KIRETT project with the focus of knowledge graphs, knowledge fusion, and user-experience-design (UX-design).

\end{abstract}


\section{Introduction}
The healthcare system is facing a paradigm shift. Demographic change and the associated shortage of doctors, nurses, and paramedics is prompting society to reconsider \cite{ kaduszkiewicz2018shortage} the way of how to treat patients efficient and optimal. Patients' desire for a data-based individual examination at the touch of a button is putting time pressure on providers to evaluate masses of data. Manually, this process can be prone to error, which could lead to incorrect treatment. Especially in rescue operations, where uncontrolled environments, language barriers, and little information about the patient's history are common issues, this can lead to major challenges in time-critical situations. With the help of modern technological approaches, such a situation can be counteracted, and the paramedic can be supported in providing patient-centered treatment. The use of modern recommendations for treatment during the operation with the integration of artificial intelligence can lead to improved treatment based on historical data and the values recorded by the paramedics on site and thus reduce late damage. In the KIRETT project, this was designed as part of a wrist-worn device \cite{zenkert_kirett_2022} and optimized in a recurring iteration with the ambulance stations in Siegen. A qualitative evaluation with fourteen participants from the rescue stations was conducted to evaluate the demonstrator for accuracy. A quantitative \cite{nadeem2024quantitative} and qualitative evaluation was conducted to measure the needs of the rescue operators with such a wearable. This paper describes the results, with the focus on knowledge graphs for treatment recommendation, UX-design for interaction-purposes of the demonstrator and knowledge fusion, of the qualitative analysis of the KIRETT demonstrator test (KDT).

\section{Background}

In the following section, the theoretical background of this paper is presented, with the focus on the project itself and the use of knowledge graphs:
\subsection{The KIRETT Project}
The KIRETT project's major goal is to assist rescue stations with an embedded wrist-worn device, in which, with the help of treatment recommendation \cite{zenkert_kirett_2022} and situation detection \cite{neuroShad}, paramedics can provide a patient-centric treatment in a time-critical situation. For that, a knowledge graph containing treatment paths \cite{nadeem2023knowledgegraph}, was developed which supports the rescue operators in their treatment with a step-by-step assistant. In addition to it, a situation detection was provided to identify treatment situations based on historical data of the rescue stations and a predefined questionnaire \cite{respiratoryShad, cardioShad}. The communication between all modules, like user interface, knowledge graph and hardware was developed. A knowledge graph was used to construct the step-by-step assistant with the help of the manual for “treatment paths and standard working procedures" \cite{treatmentpdf} used and provided by the rescue station. Further knowledge integration from n-many knowledge graph sources was considered, including Bayesian networks \cite{nadeem2024knowledgefusion} and vital signs integration \cite{nadeem2024vitals}. Various preliminary work was done for this project. Table \ref{table:related_work} presents an overview of the already published work in the KIRETT project.

\begin{table}[ht]
    \centering
    \caption{Related work in the KIRETT project }
    \label{table:related_work}
    \begin{tabular}{ccl}
    \toprule
    Topic & Source\\
    \midrule
        Overview of the KIRETT Project &  \cite{zenkert_kirett_2022} \\
        Knowledge Graph Concepts and Construction  & \cite{abu2022transferrable} \cite{frameworkAbuRasheed} \cite{nadeem2023knowledgegraph} \\
        UX-design of the KIRETT Wearable &  \cite{uxNadeem} \\
        Situation Detection &  \cite{cardioShad} \cite{respiratoryShad} \cite{neuroShad} \cite{ezekiel2023time}\\
        Knowledge Fusion & \cite{nadeem2024knowledgefusion} \\
        Vital Signs Integration & \cite{nadeem2024vitals} \\
        Quantitative Evaluation & \cite{nadeem2024quantitative} \\
        Optimization of VTA & \cite{ezekiel2023optimization} \\
    \bottomrule
    \end{tabular}
\end{table}

\newgeometry{ right=19.1 mm, left=19.1 mm, top=19.1 mm, bottom=19.1mm}

\subsection{Use of knowledge graphs in the KIRETT Project}
To achieve this goal, the help of knowledge graphs (KG) was enlisted, as they are powerful tools of knowledge representation~\cite{KRGraphs}. 
Knowledge graphs are structured repositories of information, organized in a graph-like format, capturing details about real-world entities, concepts, and their interconnections. \cite{Building_Healthcare_KG, SemanticNetworks, KRModeling, KG_Tommasini}, often modelling knowledge of a certain domain of interest \cite{What_is_KR,Ontologies}. KGs hereby find use in depicting knowledge regarding medicine domains for a varying range in applications~\cite{KRModeling}. Knowledge graphs serve as a powerful method for managing vast volumes of data, organizing it comprehensively to represent and navigate the complexities of the real world effectively. They enable the encapsulation and management of extensive information within a structured framework~\cite{KRGraphs}, and are also often used alongside machine learning and natural language processing to allow working with large and potentially unstructured datasets~\cite{KG_Tommasini}.
To build a KG in the domain of medicine to display helpful data to rescuers, medical accuracy needs to be a top priority to ensure patient safety. In the case of KIRETT, it is achieved by using the treatment manual~\cite{treatmentpdf} as the dataset it was built from, as the manual only contains approved medical procedures and treatments~\cite{zenkert_kirett_2022,frameworkAbuRasheed}.
To build the foundation for KIRETT, text mining was used to transfer the knowledge of the manual into the form of a KG to be used by the other components of the KIRETT device.

\section{Methodology}
In the following the methodology of this paper is presented, focusing on the study design and the measurements for the statistical analysis:
\subsection{Study design}
\begin{figure}[htbp]
\centerline{\includegraphics[width=\linewidth]{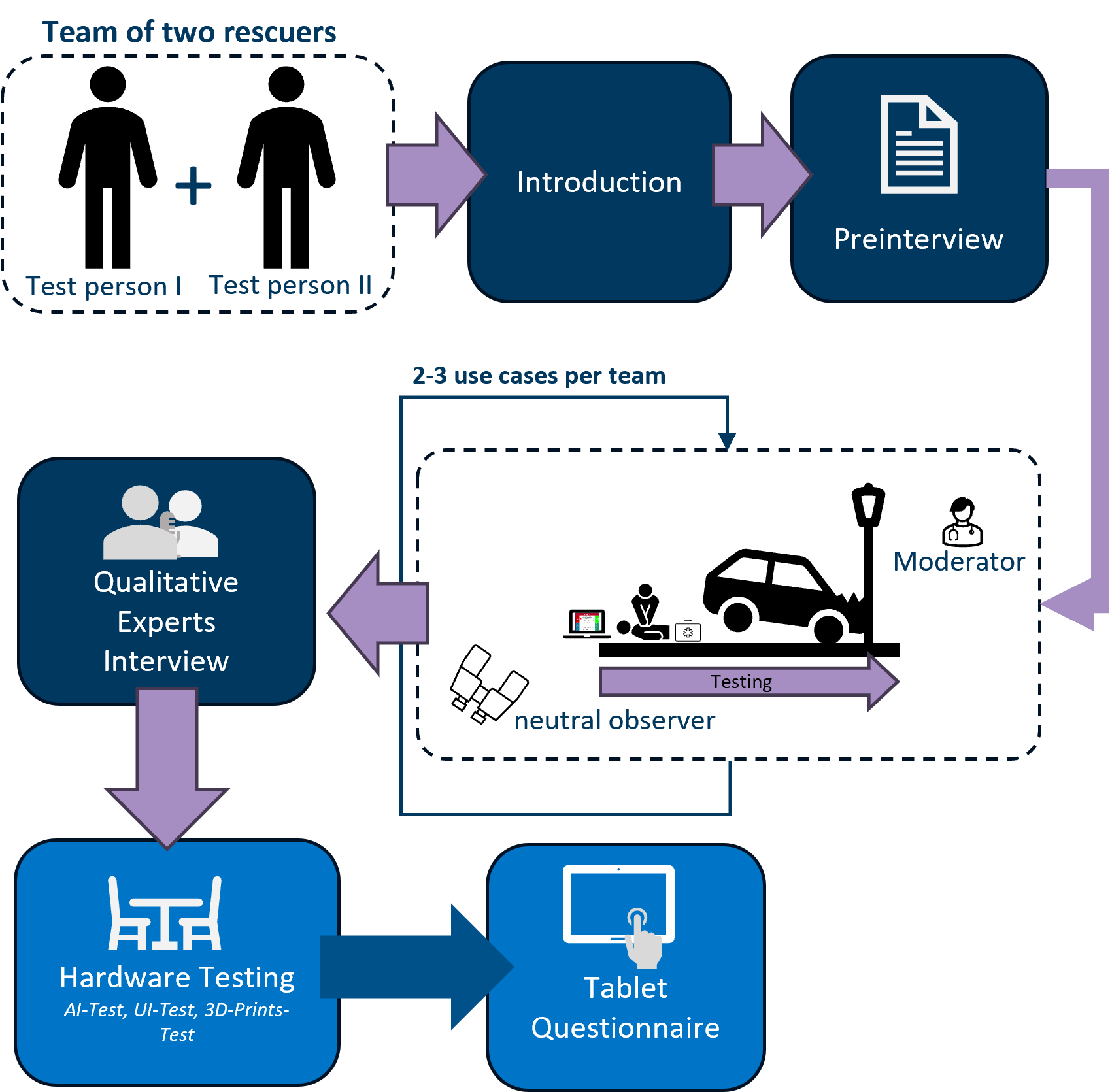}}
\caption{\textbf{Setup of the KDT: A team of two rescuers is lead through various rooms to evaluate the project. They first receive an introductory presentation with a follow-up questionnaire. Subsequently,  they perform a use case to treat a test patient in form of a medical puppet. Observations are being made on how the rescuers interact with the user interface. Afterwards, each test person is asked some questions to gain their feedback before they continue with the evaluation of  the hardware table to experience the hardware wearable and its 3D-printed case.}}
\label{generalsetup}
\end{figure}

Figure \ref{generalsetup} describes the setup of the two-day evaluation in the Firestation of Siegen-Germany. A total of fourteen testers went through  multiple scenarios interacting with the KIRETT prototype in groups of two, and underwent interviews to deliver qualitative feedback on their experiences. Even from this small number of testers, qualitative research can gather a lot of information~\cite{pope2000analysing}. However, to do so, a great amount of thought had to be put into creating the environment and setup of the evaluation~\cite{ritchie2013qualitative}, in this case to bring the rescuers into a scenario and space of mind where they can optimally use and experience the demonstrator. The KDT was therefore be split into six parts, which will be elaborated:
\subsubsection{Introduction}
\begin{figure}[ht]
    \centerline{\includegraphics[width=250
px]{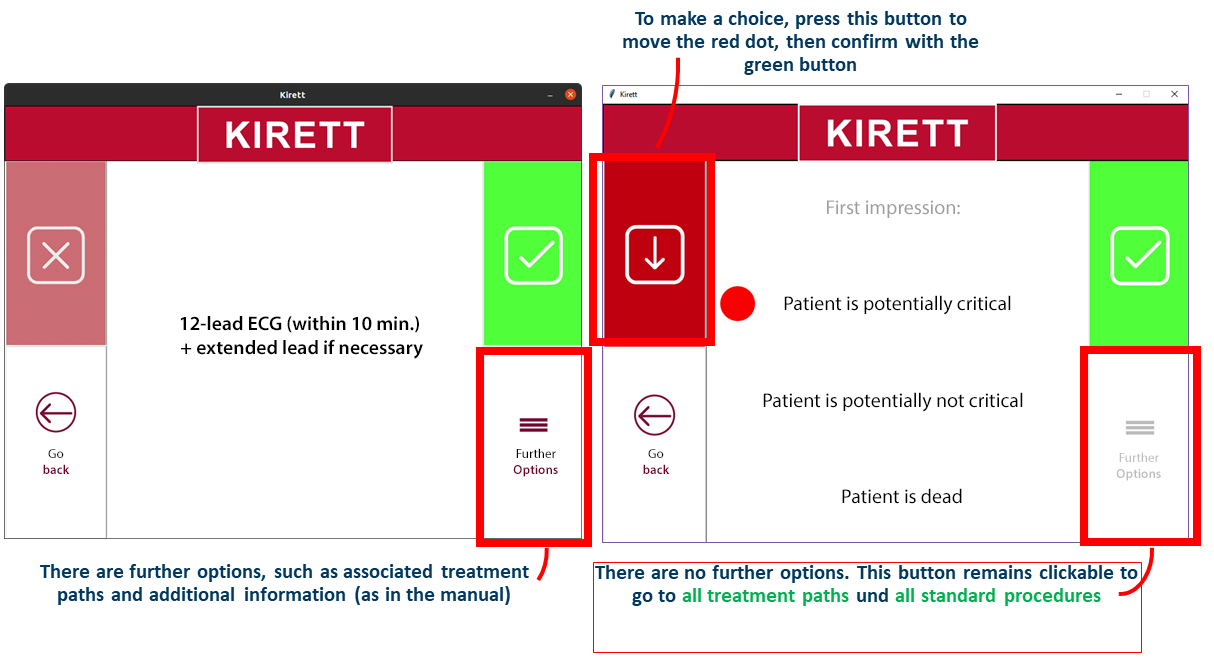}}
    \caption[Evaluation Further Options Button Availability Explanation]{
        Part of the introductory presentation of the final evaluation. It shows the functionality of the multiple-choice decision window as well as a short explanation of the `further options' button as it was used during testing.
    }
    \label{fig:Setup_Introduction_PresentationSlide}
\end{figure}

Tester coming into the KDT know little to nothing about project KIRETT, and therefore require an introduction to ensure everyone has the same pre-requisite knowledge going into the evaluation.
This presentation at the beginning of the KDT served to lay down the very foundations of what KIRETT is and how it could support rescuers.
Beyond that, it also covered the very basic interaction with the prototype by showing an introduction into the visuals and operation instruction of  the graphical user interface (GUI) (Fig. ~\ref{fig:Setup_Introduction_PresentationSlide}).

\subsubsection{Pre-interview}
\label{pre-interview}
A preinterview was conducted over a questionnaire, to gather personal information of rescuers to identify post correlations. These questions asked for the testers’ age and level of job experience to establish a baseline of testers’ life on the job. It furthermore asked questions to determine their standpoint on digitalization in rescue services and asked about the rescuer's opinion on and experiences with artificial intelligence (AI). These preliminary questions served to put the main questions in the interview after the testing into perspective. 

\subsubsection{Testing Phase}
The testing phase is the practical aspect of the KDT where rescuers are put in a realistic scenario where they can use the KIRETT demonstrator to evaluate its applications in a simulated case based on a real one.
For the prototype, a software solution was created to assess the functionality of KIRETT without relying on the resource-constrained hardware. The GUI was designed in such a way that it would be equivalent to what the hardware version could support to make for easy comparability.
In which way this GUI, which was set up on a touch tablet, would be used by the rescuers, was mostly left up to them so they could experiment with how they would use it in real cases.
Use cases were created beforehand in talks with medical experts~\cite{zenkert_kirett_2022}. They were created based on anonymized real data to make for scenarios close to the rescuers’ daily reality, and they were told to behave as though this was a real case with the addition of including the KIRETT prototype. The use cases can be viewed in table ~\ref{tab:use cases}.
The patient was voiced, and their status was given by a moderator to create a realistic environment. Each group would go through two use cases, the emphasis was put on the acute coronary syndrome (ACS) use case, which each group went through while the second use case was randomized from six pre-determined ones covering various diseases and medical problems.

\begin{table}[ht]
    \centering
    \caption{An overview of all use cases used during evaluation.}
    \begin{tabular}{ccl}
        \toprule
        Use Case Description & Diagnosis   \\
        \midrule
        Chest pain, recovering from sickness & ACS/dyspnoe \\
        Empty bottles, farewell letter, fell onto heater & Suicide attempt\\
        Chest pain, otherwise stable & Unclear chest pain \\
        Alcoholised at supermarket & Intoxication \\
        Stomach pain, nausea & Cyst \\
        Stomach ache, nausea and vomiting & Influenza\\
        \bottomrule
    \end{tabular}
    \label{tab:use cases}
\end{table}

Going through these cases allowed the rescuers to explore the KIRETT prototype in scenarios they face every day and gain their own experiences on it. When stuck on how to interact with the device, they would get hints from the team there, but largely they were left to make their own experiences with it.
During testing, impressions on their interactions were gathered by neutral observers to provide additional feedback that might otherwise have gotten overlooked in the interview step they were sent to next.

\begin{figure}[ht]
    \centerline{\includegraphics[width=200px]{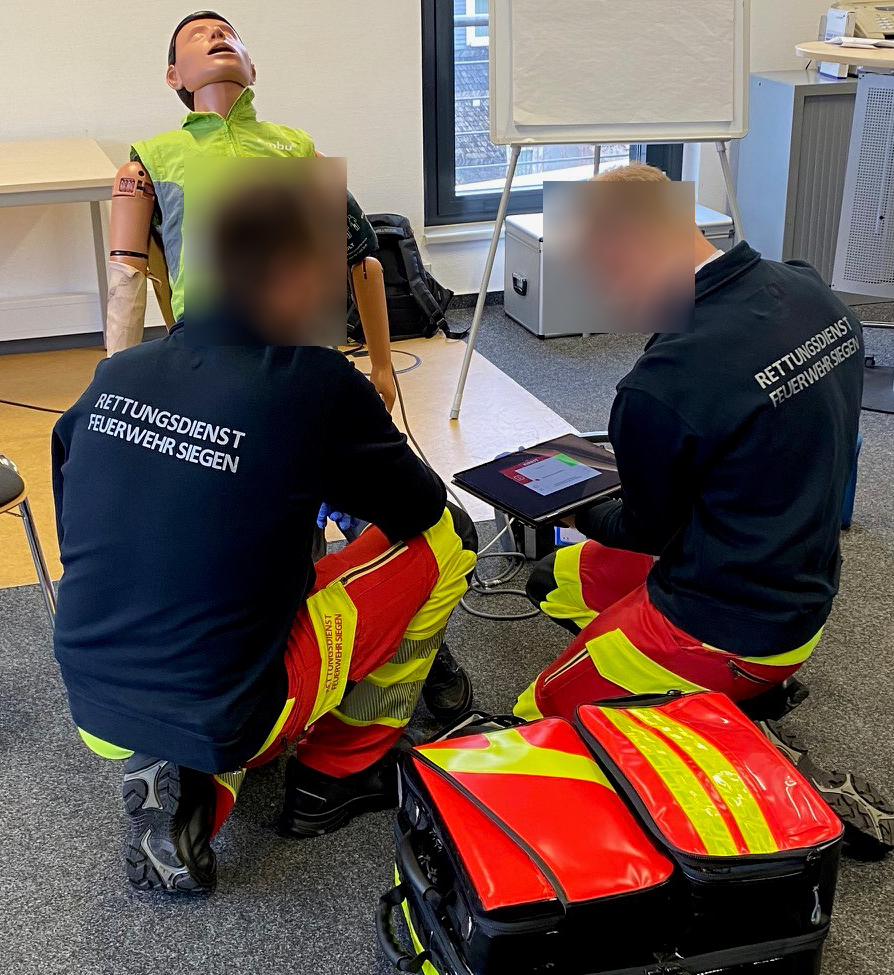}}
    \caption{
        This figure presents the dummy with the software-based KIRETT-demonstrator in active usecase-treatment with a team of rescue operators.}
    \label{fig:Setup_Testing}
\end{figure}

\subsubsection{Qualitative Expert Interviews}
To gain qualitative feedback on their interaction with the prototype, testers were interviewed after going through the use cases. The interview contained open questions for all different modules KIRETT consists of to cover a wide range of feedback to evaluate for all partners.
For the interviews, the tester groups were split up to gain each individual's unbiased opinions on the thirteen questions.
The interviews were done in a manner described by Ritchie et al. as open talks with a certain purpose to gain information on topics at hand~\cite{ritchie2013qualitative}, and asked to them in a very open manner to allow for them to give any feedback they might have.
The evaluation of the results of the interview will be done in section~\ref{InterviewEval}.

\subsubsection{Hardware Testing}
To conclude the testers’ experience with KIRETT, they were shown the hardware the device would usually run on to form an opinion on how the software solution they encountered, would feel when put into perspective on the smaller hardware screen. They were able to check out the 3D-printed cases for both the wrist-worn prototype, and the power bank and board worn around the hip and were encouraged to try them on themselves. They were able to operate a short demonstration on the hardware prototype like how they worked with the software solution and could form their opinions further while doing so.

\subsubsection{Tablet Questionnaire}
After having experienced the hardware themselves, testers were given another tablet questionnaire to answer a few questions on their opinions on the hardware. This was done similarly to the pre-interview mentioned in section~\ref{pre-interview}. 
The questions focused on the usability of the prototype in real situations, querying its runtime, weight and usability.

\subsection{Statistical Analysis}
\label{statisticalEval}
The, in this paper presented results are conducted through interview sessions with rescue operators, after a 60 to 90-minute testing phase with the demonstrator. The interviews were conducted with RODE WIRELESS GO II microphones in German and were then transcribed with MAXQDA. For this paper, statements of the Interview were translated into English. A total of fourteen participants were interviewed, which resulted in a total of 6 hours 32 minutes of qualitative recordings of the interview. As Hardware et al. \cite{hardware2015patients} presented in their paper, the characteristics of the participants are presented here \ref{table:partichari}:

\begin{table} [ht]
    \centering
    \caption{Participants characteristics \cite{nadeem2024quantitative}}
    \label{table:partichari}
    \begin{tabular}{ccl}
    \toprule
    Characteristics & Number\\
    \midrule
       Participants & 14 \\
       Mean age & 32 \\
    \midrule
    Experience level \\
        \midrule
        In training &	4\\
        0-2 years work experience&	2\\
        3-5 years of work experience	&2\\
        6-10 years of work experience&	2\\
        11+ years of work experience&	4\\
        \midrule
        Relevance of introduction of AI in rescue services \\
        \midrule
      very unimportant&		0\\
unimportant&		0\\
important&		8\\
very important&		6\\

    \bottomrule
    \end{tabular}
\end{table}

\section{Results}
\label{InterviewEval}

The following results were prepared based on expert interviews (14 in total), identifying the usability of the projects' knowledge graph, the concept of the UX design, the integration of vital signs and Bayesian networks, and the 3D-printed cases. The goal of this qualitative analysis is to gain insights of the usability, accuracy of the treatment recommendation and to gather information about improvements. This section will cover the results from the expert interviews:
In total fourteen participants were interviewed for approximately 35 minutes per participant. The participants are rescue operators from two rescue stations in Siegen (Firefighters Siegen and German Red Cross Siegen) and are between the age range of 25-34 years. Few (4 out of 14) were older, and two were younger.

\subsection{Treatment graphs' accuracy, comprehensibility, and plausibility}
In response to the interview questions, regarding the accuracy, comprehensibility and plausibility of the treatment graph provided in the KIRETT project, the majority of the test-participants confirmed, that the treatment selected and tested in this testing-phases (see table \ref{tab:use cases}) are considered to be the same as in the treatment manual (treatment manual: \cite{treatmentpdf}). They confirmed that the content of the knowledge graph closely mirrored the manual, reflecting a high medical treatment accuracy. However, for the comprehensibility and the utility of the treatment paths, the opinions differed. While some participants found the step-by-step treatment steps visualization beneficial, some testers stated the need for a more expansive view of the treatment steps, to ensure contextual understanding. It was underscored that although the treatment paths were constructed through theoretical procedures, practical application might deviate based to unique case, potentially limiting the paths' universal applicability. On several occasions, treatment steps were skipped by the rescuers in their heads during treatment, but the treatment step was displayed until the goal of the treatment step was reached. Such a visual fixation might support rescue operators as a reminder to fulfill specific tasks, which can also lead to additional time investment. A participant, who serves as a trainer for a rescue operator agreed and stated:

\begin{quote}
"Diagnostic thinking needs to go past the treatment paths from
the manual."
\end{quote}

\noindent This shows clearly that even with a rescue operation tool, such as KIRETT, diagnostic thinking needs to be at the forefront to provide the best possible treatment. 

\subsection{Visuals and integration of vital signs}
Participants were asked, to evaluate the reliability of information, like vital signs, on the display. Testers mostly agreed that the values shown during treatment with KIRETT are trustworthy. However, there were doubts about data from external sources like the control station, with concerns about its accuracy. Participants worried that if this part of the system failed, it could either stop showing values altogether or display incorrect ones.
Another issue raised was that those certain values, like temperature, aren't always automatically sent to devices. This means the KIRETT wearable might not always have all the data it needs to display accurately, relying heavily on the AI to fill in the gaps.

\subsection{Integration of Bayesian Networks based on historical data}
The integration of Bayesian network-based recommendations from historical data was also evaluated by the participants. When asked about the potential of implementing Bayesian networks for making suggestions based on historical data in the KIRETT wearable, and whether rescuers would trust the data provided by it, all respondents expressed interest and acknowledged its potential utility. While some recounted scenarios where such functionality could have been beneficial, others were cautious, indicating they would view it as a useful proposition but not rely on it entirely. One tester noted the possibility of introducing bias with this feature, while another suggested it could counteract biases towards particular treatment paths and offer alternative insights.

Regarding the graphical user interface (GUI) questions, they included inquiries about the potential implementation of future features akin to the Graph questions, as well as general feedback on whether the current interface suited users and any desired changes.

\subsection{Voice control}
The KIRETT demonstrator evaluated in this testing was built for touch input, however, the use of control via voice was evaluated with the participants. Seven participants of the evaluation (50\% of test users) expressed favor for this feature but also highlighted significant challenges:
\begin{itemize}
    \item Rescuers must communicate with team members and patients amid background noise and bystander interference.
    \item Voice recognition technology needs to distinguish relevant commands from ambient noise.
    \item Patients may feel unsettled or worried when a wearable device discusses their medical status.
    \item Encrypted exchanges among rescuers to manage patient perception of severity may not align with AI-assisted systems requiring access to such information.
    \item Concerns about maintaining a human touch in caregiving despite technological advancements.
    \item Elderly patients with hearing impairments or limited familiarity with technology may find such interactions disorienting.
\end{itemize}
Many testers concluded that having both voice controls and touch available would be beneficial, allowing rescuers to choose the most suitable method depending on the situation and the patient's needs.

\subsection{UX design of the Tkinter Environment}
The UX design was also discussed with the participants of the interview. Initially, the intuitively, clarity and simplicity of the design were evaluated. Half of the participants expressed positive views regarding the UX design of the wearable device and mentioned a good learning curve. They reported that they quickly grasped how to interact with it during initial use and found it increasingly intuitive with continued use. The simplicity of the design was appreciated, and the testers found the amount of text and size of buttons to be suitable, considering the eventual transition to a smaller, wrist-worn device. One tester stated clearly:

\begin{quote}
"More elements on the interface wouldn’t be any good, the simplicity is
great here. This tool needs to be easy to operate, even at 3am at night."
\end{quote}

\noindent However, improvements of the design were mentioned, regarding the intuitively of navigation menu, smartphone-like scroll-navigations, and concerns about the colors used in the UX-design. This was due to accessibility concerns in for example the case of a colorblind rescue operator. A participant stated that the UX-design may be more accepted by younger rescuers, but may face difficulties in acceptance by older personal, who are not as familiar with digital technologies. 

Regarding the question of further improvements of the interface, the following points were mentioned:

\begin{itemize}
    \item Integrate and consolidate treatment of children, addressing unique needs that differ from adult treatment paths and are spread across multiple apps and guidelines. 
    \item Implementation of a timer feature to aid in time-sensitive situations during rescue missions.
    \item Add haptic or visual feedback to enhance user interaction and confirm button presses.
    \item Include automatic calculation of medication dosages and documentation of medication administration and steps taken.
    \item Ensure the wearable device is easily disinfectable and compatible with various environmental conditions, including bright sunlight, nighttime, and social gatherings.
    \item Improve readability of steps and indications in all situations, including adjusting text font size for readability by rescuers of advanced ages.
\end{itemize}

\section{Discussion}

Various aspects were examined in more detail during the interviews. These will be summarized and discussed in this section.
Testers generally confirmed the accuracy of the demonstrator in reproducing the treatment manual, providing valuable step-by-step guidance. Concerns were raised that a more comprehensive representation of treatment steps was required to ensure contextual understanding, indicating potential limitations in real-world application scenarios.
While testers expressed interest in integrating Bayesian networks for treatment suggestions based on historical data, they were cautious about its reliability and potential bias. Further research is needed to address these concerns and optimize the utility of the feature.

Feedback on the additional dialogue system highlighted its potential benefits, but also outlined significant challenges such as background noise. In addition, suggestions were made to improve the design of the user interface, including the intuitiveness of the navigation menu and consideration of accessibility settings for colorblind users. Further suggestions for the user interface and functionality included integrating treatment for children, adding timer functions, providing haptic or visual feedback, and improving readability for users of all ages.
Considering the feedback and recommendations from the qualitative analysis could significantly improve the usability, effectiveness and acceptance of the demonstrator by rescue workers in different operational environments. Future iterations should prioritize these findings to optimize the demonstrator for real-life rescue scenarios. 

\section{Limitations}
Despite the valuable insights gained from the qualitative analysis of the demonstrator evaluation, it is important to acknowledge certain limitations associated with the study design and methodology.

The study was conducted with a relatively small sample size of 14 testers, which may limit the generalizability of the results. In addition, the composition of the tester group, which consisted mainly of ambulance staff from specific stations in Siegen, Germany, may not fully represent the diverse perspectives and experiences within the wider ambulance community.

The evaluation was conducted in a controlled environment at the fire station in Siegen-Germany. While efforts were made to simulate realistic rescue scenarios, the artificial nature of the test environment may not fully reflect the complexities and challenges encountered in actual rescue operations.

The qualitative analysis relied on subjective feedback obtained through interviews and (neutral) observations. While this approach provided a comprehensive and in-depth insight into the testers' experiences and perceptions, it also carries the risk of bias and misinterpretation. Interviews were conducted in German and then translated for the analysis.

The length of the evaluation sessions, which were typically 60 to 90 minutes, may have limited the depth and breadth of feedback received from the testers. Longer observation periods or follow-up sessions could have provided additional insight into the usability and effectiveness of the demonstrator over time.

The demonstrator itself may have inherent limitations in terms of its functionality, reliability, and compatibility. These technological limitations could impact the overall user experience and effectiveness of the demonstrator in real-world operational scenarios.

\section{Conclusion}
Despite the limitations, the results of the qualitative analysis provide valuable insight and guidance for future research and development efforts aimed at improving the usability, reliability and practicality of the KIRETT demonstrator in supporting rescue operations. Furthermore, the qualitative results provide valuable guidance for refining the KIRETT demonstrator and maximizing its potential impact on improving rescue services.  
KIRETT's advantages in supporting rescuers in their work as well as some problems that currently pose disadvantages, were identified in this evaluation, and could be addressed with further research.
Continued collaboration with end users and stakeholders will be crucial for the further development and implementation of this innovative technology.

\section*{Funding}
This ongoing research is funded by the German Federal Ministry of Education and Research (BMBF). Futhermore it is coordinated by CRS Medical GmbH (Aßlar, Germany) with support from partner mbeder GmbH (Siegen, Germany). We also would like to present our sincere gratitude to associative partners: Kreis Siegen-Wittgenstein, City of Siegen, German Red Cross Siegen (DRK), and Jung-Stilling-Hospital in Siegen, Germany.

\bibliographystyle{unsrt}  
\bibliography{templateArXiv}

\end{document}